\documentclass[usenatbib]{mn2e}

\usepackage{graphicx}

\title[The HeII$\lambda 4686$ line profile in $\eta$ Carinae]
{Wind-wind collision in the $\eta$ Carinae binary system - III. The He\,{\Large\bf  II} $\lambda 4686$ line profile}
\author[Z. Abraham and D. Falceta-Gon\c{c}alves]
{Z. Abraham$^{1}$\thanks{E-mail:zulema@astro.iag.usp.br}, D. Falceta-Gon\c{c}alves$^{1,2}$ \\
$^{1}$Instituto de Astronomia, Geof\'\i sica e Ci\^encias Atmosf\'ericas, Universidade de S\~ao Paulo, Rua do Mat\~ao 1226,\\ Cidade Universit\'aria 05508-090, S\~ao Paulo, Brazil\\
$^{2}$N\' ucleo de Astrof\' isica Te\' orica, CETEC, Universidade Cruzeiro do
Sul, Rua Galv\~ ao Bueno 868, CEP 01506-000 S\~ao Paulo, Brazil\\}
             
\begin{document}

\date{ }

\pagerange{\pageref{firstpage}--\pageref{lastpage}} \pubyear{2007}

\maketitle

\label{firstpage}

\begin{abstract}

We modeled the He\,{\sevensize II} $\lambda 4686$ line profiles  observed in the $\eta$ Carinae binary system close to the 2003.5 spectroscopic event, assuming that they were formed in the shocked gas that flows at both sides of the  contact surface formed by wind-wind collision. We used a constant flow velocity and added turbulence in the form of a gaussian velocity distribution. We allowed emission from both the primary and secondary shocks but introduced infinite opacity at the contact surface, implying that only the side of the contact cone visible to the observer contributed to the line profile. Using the orbital parameters of the binary system derived from the 7 mm light curve during the last spectroscopic event (Paper II) we were able to reproduce the line profiles obtained with the $HST$ at different  epochs, as well as the line mean velocities obtained with ground based telescopes. A very important feature of our model is that the line profile  depends on the  inclination of the orbital plane; we found that to explain the latitude dependent mean velocity of the line, scattered into the line of sight by the Homunculus, the orbit  cannot lie in the Homunculus equatorial plane, as usually  assumed. This  result, together with the relative position  of the stars during the spectroscopic events, allowed us to explain most of the observational features, like the variation of the ``Purple Haze'' with the orbital phase, and to conciliate the X-ray absorption with the postulated shell effect used to explain the optical and UV light curves.

\end{abstract}

      
\section{Introduction}

$\eta$ Carinae is a massive luminous blue variable (LBV) with a high mass loss rate, which suffered several episodes of large mass loss (Davidson \& Humphreys 1997). 
It  presents also periodic spectroscopic events that led to the fading of the high excitation lines and a decrease in the continuum emission (Damineli 1996). 
The last event, predicted  for epoch 2003.5, was duly observed at wavelengths ranging from radio to X-rays, through intense follow up campaigns (Fern\'andez Laj\'us et al. 2003,  van Genderen et al. 2003, Groh \& Damineli 2004, Whitelock et al. 2004, Abraham et al. 2005a, Davidson et al. 2005,  Weis et al. 2005, Nielsen, Gull \& Vieira Kober 2005, Corcoran 2005, Hartman et al. 2005, Martin et al. 2006, Gull, Vieira Kober \& Nielsen 2006).

Although it does not remain any doubt about the strict periodicity  of the spectroscopic events, the 2003.5 occurrence  did not help elucidate their true nature.  Moreover, the high quality data allowed the discovery of new features, like (i) a sharp peak in the 7 mm radio emission, coincident with the spectroscopic event and superposed to the general weakening trend (Abraham et al. 2005a), (ii) the identification of the He\,{\sevensize II} $\lambda 4686$ recombination line, which reached very large intensity before disappearing  (Steiner \& Damineli 2004), and (iii)  the latitude dependence of this emission (Stahl et al. 2005).

The early models proposed to explain the spectroscopic events  postulated the ejection of shells from the stellar surface (Zanella, Wolf \& Stahl 1984) but the periodicity reported by Damineli (1996) in the intensity of the He\,{\sevensize I} $\lambda$10830 line as well as in the radial velocity of the Pa$\gamma$ and Pa$\delta$  lines gave strong support to the existence of a binary companion to $\eta$ Carinae, moving in a highly eccentric orbit (Damineli, Conti \& Lopes 1997, Davidson 1997, Damineli et al. 2000). The nature of the secondary star, assumed to be a WR with a strong wind, was recently confirmed  by Iping et al. (2005) from data obtained with the {\it Far Ultraviolet Spectroscopic Explorer (FUSE)} satellite.  

The period of the binary system is very well determined but the other orbital parameters are controversial. The main reason is that the stellar  photosphere, from which the orbital radial velocities  are usually determined, is not visible in any of the stars. Assuming that the radial velocity of the Paschen  lines reflects  the orbital motion of the primary star, the best fitting  orbital parameters resulted in eccentricities between 0.6 and 0.8, with the ellipse major axis forming a small angle with the line of sight,  and $\eta$ Carinae positioned between the observer and the companion star during periastron passage (Damineli et al. 1997, Davidson 1997). Besides, the orbital plane was always assumed to coincide with the Homunculus equator. 

This orientation was appropriate to explain the  fading of the X-ray flux during periastron passage,  which would be absorbed by the dense $\eta$ Carinae wind (Pittard et al. 1998; Corcoran et al. 2001; Pittard \& Corcoran 2002) but it  
failed to explain the disappearance of the high excitation lines (Davidson et al. 1999, Martin et al. 2004). 

Other phenomena,  like a phase dependent asymmetry in the intensity of the UV emission reflected in the Homunculus, reported by Smith et al. (2004), was  attributed to the shadow of the primary star on the light of the secondary, in a model that requires the major axis of the orbit to be perpendicular to the line of sight.  

Finally, $HST$ observations, with higher spatial resolution than those obtained by Damineli et al. (1997, 2000) with ground based telescopes, could not reproduce the expected mean velocities of the lines at some of the orbital phases (Davidson et al. 2000).

On the other hand, from a more pragmatic point of view we must consider that, although $\eta$ Carinae  is probably one of the most massive stars known at the present time, the properties of the binary system it belongs to must be similar to those of others formed by WR and O stars, like  WR$\;140$, WR$\;106$, WR$\;104$ and WR$\;137$ (Monnier, Tuthill \& Danchi 2001; Cohen \& Vogel 1978, Williams; Kidger \& van der Hucht 2001; Kato et al. 2002)

Based on these considerations,  Falceta-Gon\c{c}alves, Jatenco-Pereira \& Abraham (2005, Paper I) showed that in the $\eta$ Carinae binary system  near  periastron passage, dust can be formed and grow very fast in  the contact surface formed by wind-wind collision. This dust, as it moves away from the stars, could absorb not only the X-rays but also the UV radiation, like in a shell ejection.  For this scenario to be effective, the secondary star must be located between $\eta$ Carinae and the observer at periastron passage, implying that the broad emission line velocities do not represent the true orbital motion.   

In a similar way, other binary systems, like WR$\;79$, present  broadened emission lines with phase dependent velocities and profiles; these lines are supposedly formed in the shocked material that flows along the contact surface of the colliding winds (Seggewiss 1974; Luhrs 1997; Hill et al. 2000; Barzakos, Moffat \& Niemela 2001). Falceta-Gon\c{c}alves, Jatenco-Pereira \& Abraham (2006) were able to reproduce the line profiles and mean velocities of the C\,{\sevensize III} $\lambda$5696  lines of the Br22 system,  assuming constant flow velocity along the wind contact surface and including turbulence and finite opacity in their calculations. In the present paper we apply the same model to the He\,{\sevensize II} $\lambda$4686 line, detected in $\eta$ Carinae by Steiner \& Damineli (2004) and Martin et al. (2006) during the 2003.5 spectroscopic event, using the orbital parameters derived by  Abraham et al. (2005b, Paper II) from the 7 mm observations of  the 2003.5 spectroscopic event. In Sect. 2 we describe the line profiles obtained by Steiner \& Damineli (2004) with ground base telescope and by Martin et al. (2006) with the $HST$, as well as the lines reflected at the Homunculus observed with the European Southern Observatory (ESO) Very Large Telescope (VLT) (Stahl et al. 2005).  We also describe the geometric model from which the line profile will be calculated and its dependence on the orbital parameters. In Section 3 we present the model profiles  obtained using  the orbital parameters derived in Paper II, and compare them with the observations. We  also investigated the inclination of the orbit, by fitting the Homunculus reflected line profiles. In Section 4 we discuss the implications of the orbital model on the observational properties at other wavelengths; a  summary of the results is presented in Section 5.

\section{Modeling the He\,{\sevensize II} $\lambda$4686 line}
\subsection{Observational data}

Steiner \& Damineli (2004) identified the He\,{\sevensize II} $\lambda$4686 line in the spectra of $\eta$ Carinae at epochs near the 2003.5 spectroscopic event. They found that the line profile changed with the phase of the secondary orbit and that its mean velocity varied systematically between -400 km s$^{-1}$ and -100 km s$^{-1}$, reaching  positive values near periastron passage. The light curve was similar to that of X-rays, with the equivalent width becoming very large before fading. The authors argued  that the He\,{\sevensize II} present in the dense stellar winds could be ionized by  the UV and soft X-rays emitted by the shocked material in the wind-wind collision region, producing the He\,{\sevensize II} $\lambda$4686 line during recombination.

The line was also detected in high spatial resolution $HST$ observations by Martin et al. (2006), who claimed that the energy in the form of soft X-rays and UV radiation is not  enough to produce, by photoionization, the amount of He$^{++}$ necessary to obtain the $\lambda$4686 line luminosity. They suggested instead that ionization could be produced by collisions in the shocked material itself, the shock being either produced by wind-wind collision or by matter ejected from the primary star at the low excitation phase.

A completely different model was proposed by Soker (2001, 2004). He assumed  that in the binary system, the secondary accretes matter from $\eta$ Carinae; in this scenario, the Homunculus was formed in the 1840 Great Eruption by a bipolar jet originated  in the secondary star. Presently, matter would be accreted   during each periastron passage, the wind of the secondary star would not reach its terminal velocity before colliding with the primary's wind, and the stagnation   point would get closer to the secondary; it is in this region  that the He\,{\sevensize II} $\lambda$4686 line would be formed. Eventually the accretion would shut down the secondary wind, producing the spectroscopic events. This model, however, fails to  explain other phenomena, like  the ``shell-like'' effect, the recombination of the disk material that results in the fading of the radio emission, and the superimposed 7 mm peak emission (Abraham et al. 2005a). 

Finally, scattering into the line of sight of the He\,{\sevensize II} $\lambda$4686 line photons by the Homunculus suggests that the emission is latitude dependent. In fact, there is a location in the nebula, labeled FOS4, which gives a reflected pole-on view of the stellar wind, with a time delay of about 20 days relative to the direct view (Meaburn, Wolstencroft \& Walsh 1987, Davidson et al 2001, Smith et al. 2003). Observations with the VLT Ultraviolet-Visual Echelle Spectrograph (UVES) showed line profiles different from those obtained by direct observation, even when corrected by travel time and Doppler effects due to the expansion of the Homunculus  (Stahl et al. 2005). None of the models proposed up to the present addresses this issue.
 
\subsection{The line profile model}

In this paper we will assume that the He\,{\sevensize II} $\lambda 4686$ line is formed in the shocked material, as also pointed out by Martin et al. (2006). This scenario was already proposed to explain the variation with orbital phase of other broadened line profiles in several  binary systems formed by Wolf-Rayet and O-type stars (eg. Seggewiss 1974, Luhrs 1997). 

Our model  assumes that the shocked material   flows  with constant velocity at both sides of the conic  contact surface, where the  momenta of the two winds balance each other.  The high temperature plasma ($10^6- 10^8$ K) emits at X-ray energies;  when the wind density is high,  radiative losses cool down the gas generating a range of  temperatures and densities, favoring the production of turbulence.  Lines can be formed under such conditions; their profile can be calculated from the projection of the flow velocity in the direction of the line of sight, which is different at different orbital phases. 

Falceta-Gon\c{c}alves, Abraham \& Jatenco-Pereira (2006) have discussed extensively this dependence; in the absence of absorption,  the line profile presents two peaks for most phases, but  the blueshifted peak dominates when the emission of the receding part of the flow is absorbed.
In $\eta$ Carinae, because of the highly eccentric orbit, the density of the shocked material becomes very high near periastron passage, radiative cooling is very fast and grain formation can occur at the contact surface, justifying the inclusion of absorption in the calculation of the line profiles. 

The contact surface geometry is described by the equation:

\begin{equation}
\medskip
\frac{dy}{dz}= \frac {(\eta^{-1/2}{d_2}^2+ {d_1}^2)y}{\eta^{-1/2}{d_2}^2z+{d_1}^2(z-D)},
\medskip
\end{equation}

\noindent
where $D$ is the distance between the  stars; $d_1$ and $d_2$ are 
the distances of the primary and secondary stars to the 
contact surface, respectively, and $\eta=\dot{M_s} v_s/\dot{M_p} v_p$, 
where  $\dot{M_p}$ and $\dot{M_s}$ are the mass
 loss rates of primary and the secondary stars, and
$v_p$ and $v_s$ their respective wind velocities. 

Asymptotically, this surface can be approximated by a cone with an opening angle $\beta$, which depends only on the parameter $\eta$ (Luo, McCray \& Mac-Low 1990; Stevens, Blondin \& Pollock).

The  flow velocity $v_{\rm flow}$ of the shocked winds, projected into the line of sight $v_{\rm obs}$ is given by:

\begin{eqnarray}
\medskip
v_{\rm obs} = v_{\rm flow} (-\cos \beta \cos \varphi\sin i + \sin \beta \cos 
\alpha \sin \varphi \sin i \nonumber \\
- \sin \beta \sin \alpha \cos i),
\medskip
\end{eqnarray} 

\noindent
where $i$ is the inclination of the orbital plane  with respect to the line 
of sight, $\alpha$ is the cone azimuthal angle and $\varphi$ is the   angle between the cone axis and the projection of the line of sight on the orbital plane.

To calculate the line profile, we introduced local turbulence at each emitting fluid element, as a gaussian distribution of velocities around the central velocity $v_{\rm obs}$ with a dispersion $\sigma$; the intensity as a function of the velocity $v$ can be calculated from:

\begin{equation}
\medskip
I(v) = \mathcal{C(\varphi)} \int_{0}^{\pi }\exp \left[ -\frac{\left( v-v_{\rm obs}\right) ^{2}}{2\sigma
^{2}}\right]e^{-\tau(\alpha)} d\alpha, 
\medskip
\end{equation} 

\noindent
where $\mathcal{C(\varphi)}$ is a constant that depends on the line intensity at each phase angle $\varphi$, and $\tau(\alpha)$ is the optical depth, which is zero for the elements in front of the contact surface  and has a constant value $\tau_0$ for the emission produced behind it. 

Therefore, besides the orbital parameters, three quantities  are needed to reproduce the He\,{\sevensize II} $\lambda$4686 line profiles  obtained by Steiner and Damineli (2004), and Martin et al. (2006): the flow velocity $v_{\rm flow}$, the width of the gaussian turbulence distribution $\sigma$ and the optical depth $\tau_0$. 
\begin{figure}
\centering
\includegraphics[width=8cm]{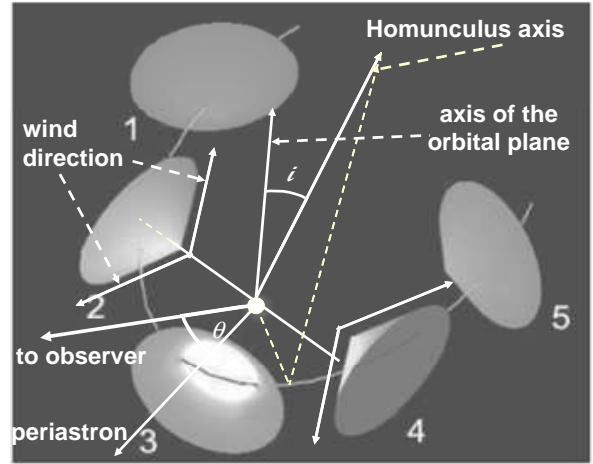}
\caption {Schematic view of the $\eta$ Carinae binary system. The wind-wind collision contact surface is shown at different positions in the orbit. The angle $\iota$ is the relative orientation of the orbital plane to the axis of the Homunculus. The angle $\theta$ is the angle between periastron and the projection of the line of sight into the orbital plane. In positions 1 and 5 of the orbit the observer will see the emission from the primary wind, with positive velocity, in position 2 and 4, both positive and negative velocities will be present, while in position 3 only the wind of the secondary star will be observed, presenting two peaks}.
\label{figure1}
\end{figure}

\subsection{Orbital parameters}

A schematic view of the wind-wind contact surface along the orbit of the binary system and its orientation relative to the observer and to the Homunculus axis can be seen in Figure 1.
To calculate the line profiles we  used the orbital parameters  derived  from 7 and 1.3 mm observations during the 2003.5 spectroscopic event (Paper II). At that epoch, the observed millimeter light curves were formed by the superposition of two components: the fading emission of an optically thick disk that surrounds the binary system, and a sharp peak which we interpreted as free-free emission from the  shock surface. We assume that the plane of the binary orbit coincides with the plane of the ionized disk, and unlike all other models, it is seen almost  edge-on ($i=90\degr$); this scenario was already proposed by Duncan \& White (2003), based in the results of the 6 cm interferometric observations. For an asymptotic  cone opening angle $\beta = 56\degr$, which  corresponds to $\eta = 0.2$ (Paper II), the best fitting of the millimeter wave light curve was obtained for an orbit with eccentricity $e=0.95$,  phase angle between conjunction and periastron passage  $\theta = 40\degr $, epoch of conjunction 2003 June 29, and periastron occurring 1.5 days after conjunction. 
\begin{table}
\caption{Parameters of the  He\,{\sevensize II} $\lambda 4686$ line at different epochs}
\begin{tabular}{@{}lcccl}
\hline
Epoch &  phase & $\varphi$ & $D$ & $C(\varphi)$ \\
JD-2,400,000 &  &degrees   & AU &  \\
\hline
52727.3 & 0.954 & -108  & 9.4 & 0.0010 \\
52764.2 & 0.972 & -101  & 6.8 & 0.0021 \\
52778.5 & 0.979 & -96   & 5.6 & 0.0022 \\
52791.7 & 0.986 & -89   & 4.4 & 0.0029 \\
52813.7 & 0.997 & -56   & 1.7 & 0.0042\\

\hline
\end{tabular}
\end{table}

\begin{figure}
\centering
\includegraphics[width=8cm]{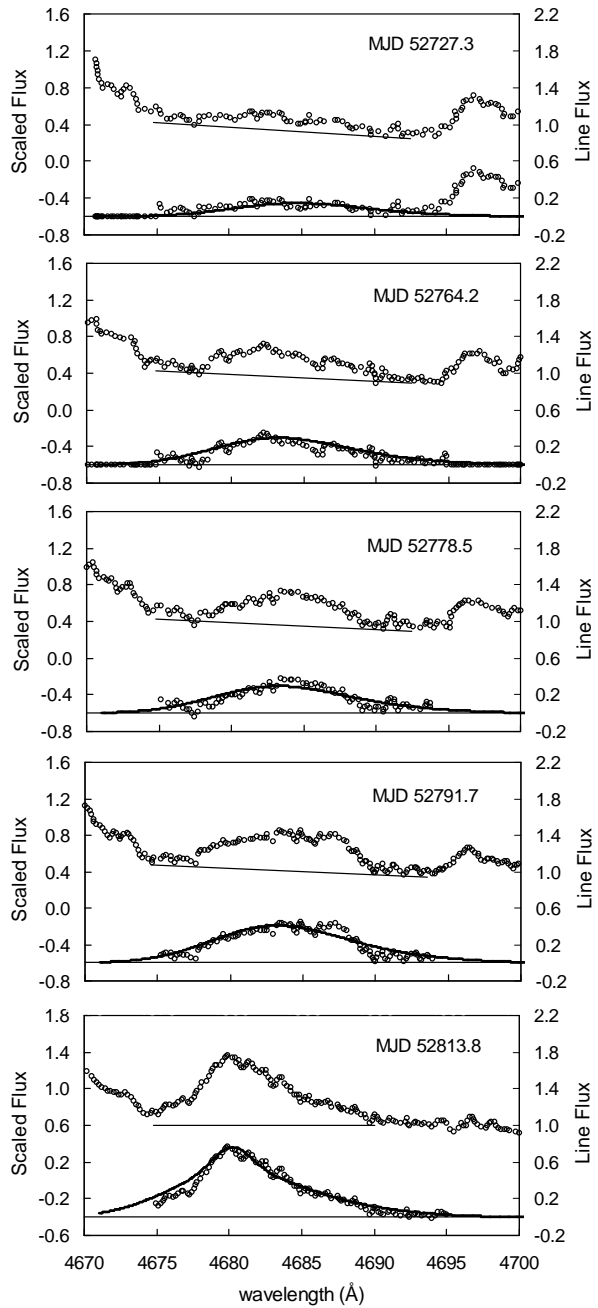}
\caption {Digitalized   line profiles, from Martin et al. (2006), and continuum baseline (open circles and straight line, respectively)  in the upper part of each panel; continuum subtracted and model line profiles (open circles and solid line, respectively) in the lower part, are shown for  five epochs near the 2003.5 spectroscopic event.}
\label{figure2}
\end{figure}

\section{Results}
\subsection{Fitting the HST  He\,{\sevensize II} $\lambda 4686$ line profiles}

To determine the flow velocity and turbulence amplitude, we fitted our model to the line profiles published by Martin et al. (2006),  since they have better spatial resolution than those obtained with a ground base telescope by Steiner \& Damineli (2004). We  assumed that the optical depth across the contact surface is $\tau _0 \gg 1$, since close to periastron passage, radiative cooling is very fast and grain formation occurs on timescales of hours (Paper I).
We used the data of the five epochs in which at least some signal was detected and we digitalized  the  profiles in order to subtract a  baseline; the resulting profiles are presented for each epoch in Figure 2 as open circles. The subtracted baselines were very close to those given by Martin et al. (2006), except for epoch MJD 52813.8, for which we used values intermediary between those proposed by Martin et al. (2006) and Steiner \& Damineli (2004); the corresponding functions are shown  as straight lines in Figure 2.

In Table 2 we present the MJD for each observation, as well as the corresponding orbital phase, obtained assuming  an orbital period  of 2022.1 days, and defining phase 1.00 as June 29, 2003 (Steiner \& Damineli 2004). Using the orbital parameters described in the previous section, we calculated for each epoch the angle between the axis of the contact cone surface and the line of sight, and the separation between the two stars, which are presented in columns 3 and 4 of Table 1, respectively. According to this model, between the first and last epochs, the angle $\varphi$ changed between $-110\degr$ and $-56\degr$, while the distance between the stars changed between 9.4 and 1.7 AU.
 
 The flow velocity and the turbulence that best fitted  the observed profiles at all epochs were $v_{\rm flow}\sin \beta =229$ km s$^{-1}$ and $\sigma = 0.33v_{\rm flow} $ km s$^{-1}$. The scaling factors $C(\varphi)$ were determined for each epoch to fit the observed line brigthness; their values are presented in the fifth column of Table 1 and the resulting model line profiles as solid lines in Figure 2, superimposed to the continuum subtracted  digitalized data. Notice that at the last epoch (MJD 52813.8), the angle between the axis of the contact surface and the line of sight was equal to the cone opening angle ($\varphi= \beta=56\degr$), meaning  that one edge of the cone was pointing towards the observer.

\begin{figure}
\centering
\includegraphics[width=8cm]{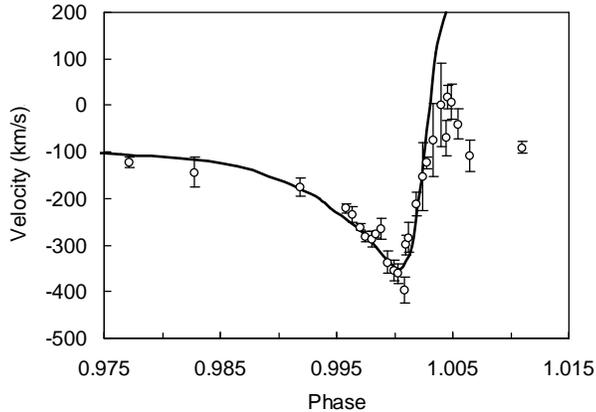}
\caption { Mean velocity of the He\,{\sevensize II} $\lambda$4686 line as a function of the orbital phase. Points are the data published by Steiner et al. 2004, the solid line represents the values obtained from our model.}
\label{figure3}
\end{figure}
\subsection{Fitting the He\,{\sevensize II} $\lambda$4686 line mean velocity}

As we can see from Figure 2 there is an excellent agreement between the model and observed profiles for all epochs in the $HST$ observations. However, the detected profiles reach up only to orbital phase 0.997, with a negative detection at phase 1.003, while the closer spaced observations of Steiner \& Damineli (2004) included many intermediary epochs. For that reason, we constructed model profiles and determined their mean velocities, for  orbital phases between 0.975 and 1.005, and compare them with the data; they are shown in Figure 3 as a solid line and open circles, respectively. The agreement between the two sets is very good up to phase 1.003, for later phases the model predicts positive velocities much larger than those reported by Steiner \& Damineli (2004). However, Martin et al. (2006) did not detect any emission from the He\,{\sevensize II} $\lambda$4686 line at phase 1.003 or later, attributing the profiles observed by Steiner  \& Damineli (2004) to  possible contamination from other nebular lines at the observed wavelengths. As it will be discussed later, absorption by the gas and dust shell formed around the system would be responsible for the absence of this line after conjunction.

\subsection{Fitting the VLT UVES He\,{\sevensize II} $\lambda 4686$ line profiles for different orbital inclinations}.

In the previous sections we showed that a very good fitting can be obtained for the He\,{\sevensize II} $\lambda 4686$ line profiles and mean line velocities using the orbital parameters derived in Paper  II from the 7 mm light curve during the 2003.5 spectroscopic event; however,  changes in the orbital inclination, from $\mid i\mid \leq 90\degr$ to $\mid i\mid \geq \beta$, could be compensated by changes in the area of the emitting cone and still explain the mm-wave observations. Here, we also found that making small changes in the flow velocity and turbulence, a satisfactory fitting to the phase dependent  profiles of the He\,{\sevensize II} $\lambda 4686$ line and their mean velocity could be obtained for other orbital inclinations. This is only a consequence of  the reduced range of orbital phase angles at which the emission was observed and modeled.

Further constrains to the orbital inclination can be obtained from the VLT UVES observations, which give the radiation emitted in the direction of the Homunculus Nebula southern pole,  scattered towards  the observer and  delayed by about 20 days from the direct view because of the light travel time (Stahl et al. 2005). Martin et al. (2006) presented these  profiles for  the epochs which coincided with the HST observations, with wavelengths already corrected for the Homunculus expansion velocity; they pointed out  the strong latitude dependence, both in the profile shapes and in their mean velocity.

\begin{figure}
\centering
\includegraphics[width=8cm]{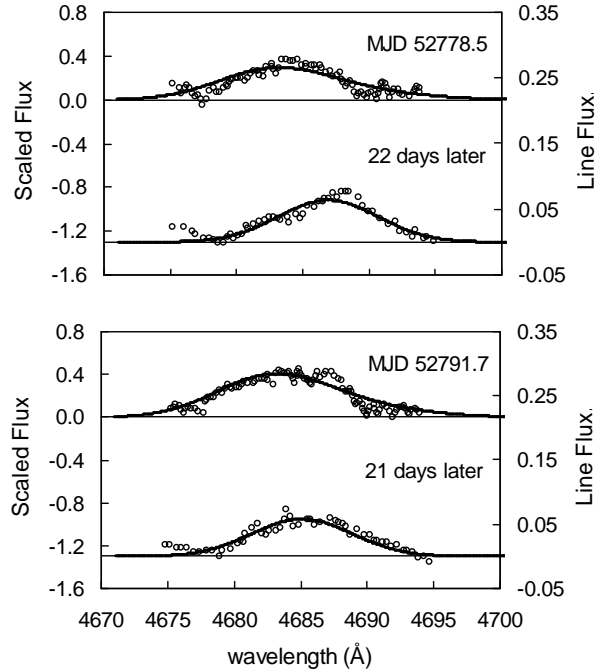}
\caption {Continuum subtracted He\,{\sevensize II} $\lambda$4686 line profiles for two epochs; upper part of the panels corresponds to the direct view and the lower part to the radiation   scattered  by the Homunculus Nebula into the line of sight. Open circles are the digitalized  data from Martin et al. (2006) and the solid lines the model profiles, obtained with inclination $i=90\degr$ for the direct view and $i^*=59\degr$ for the reflected radiation.}
\label{figure4}
\end{figure}

\begin{figure*}
\centering
\includegraphics[width=13 cm]{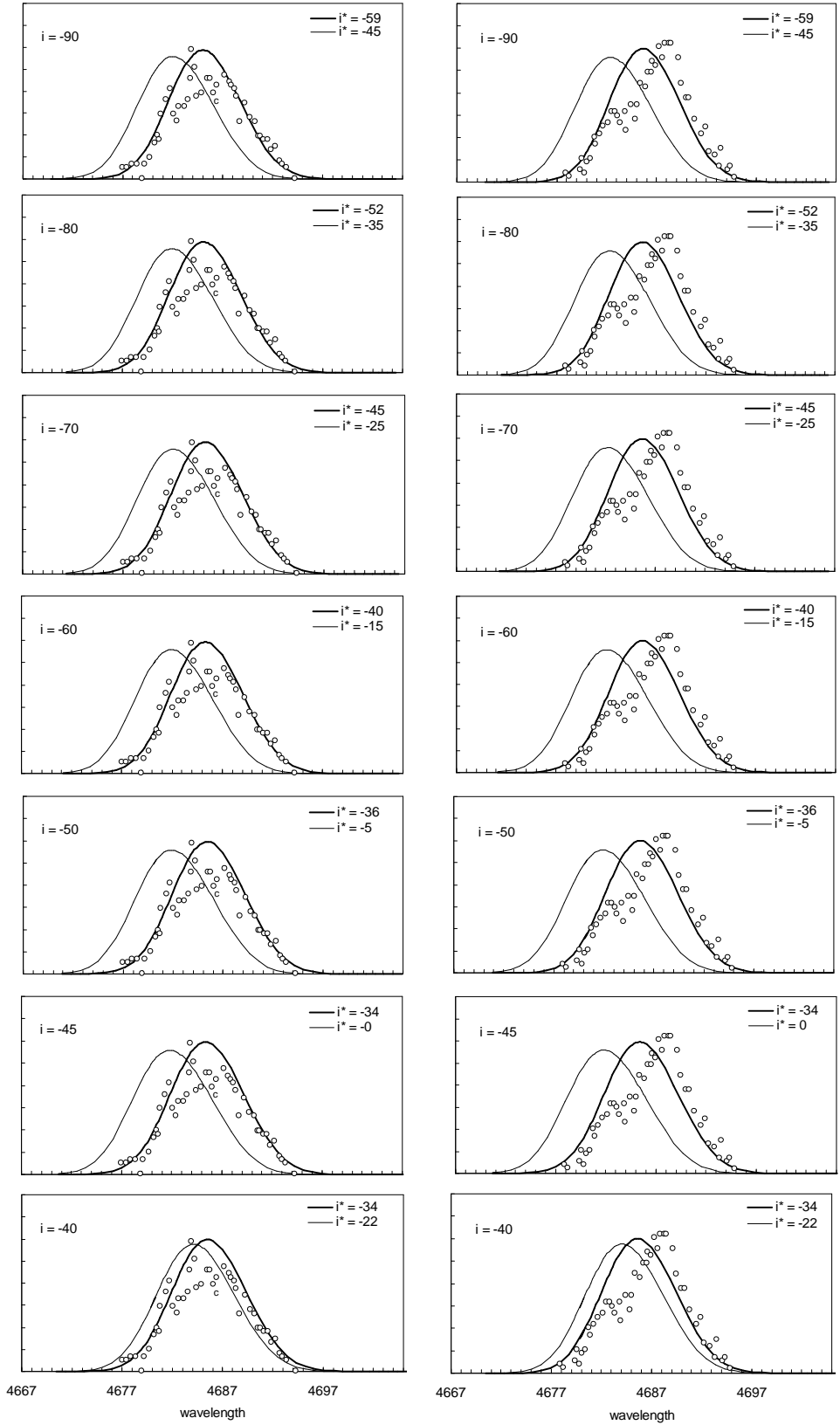}
\caption {Left and right panels: same  He\,{\sevensize II} $\lambda$4686 line profiles reflected by the Homunculus as in Figure 4 (points), and model profiles for  inclinations  $i^* = 90\degr, 80\degr, 70\degr, 60\degr, 50\degr, 45\degr$ and $40\degr$ that best fitted the observations (heavy  line), and minimum $i^*$ compatible with equation 4 (light  line). Values of $i^*$ smaller than those shown in the graph would produce double peaked profiles.  }
\label{figure5}
\end{figure*}

\begin{figure*}
\centering
\includegraphics[width=13 cm]{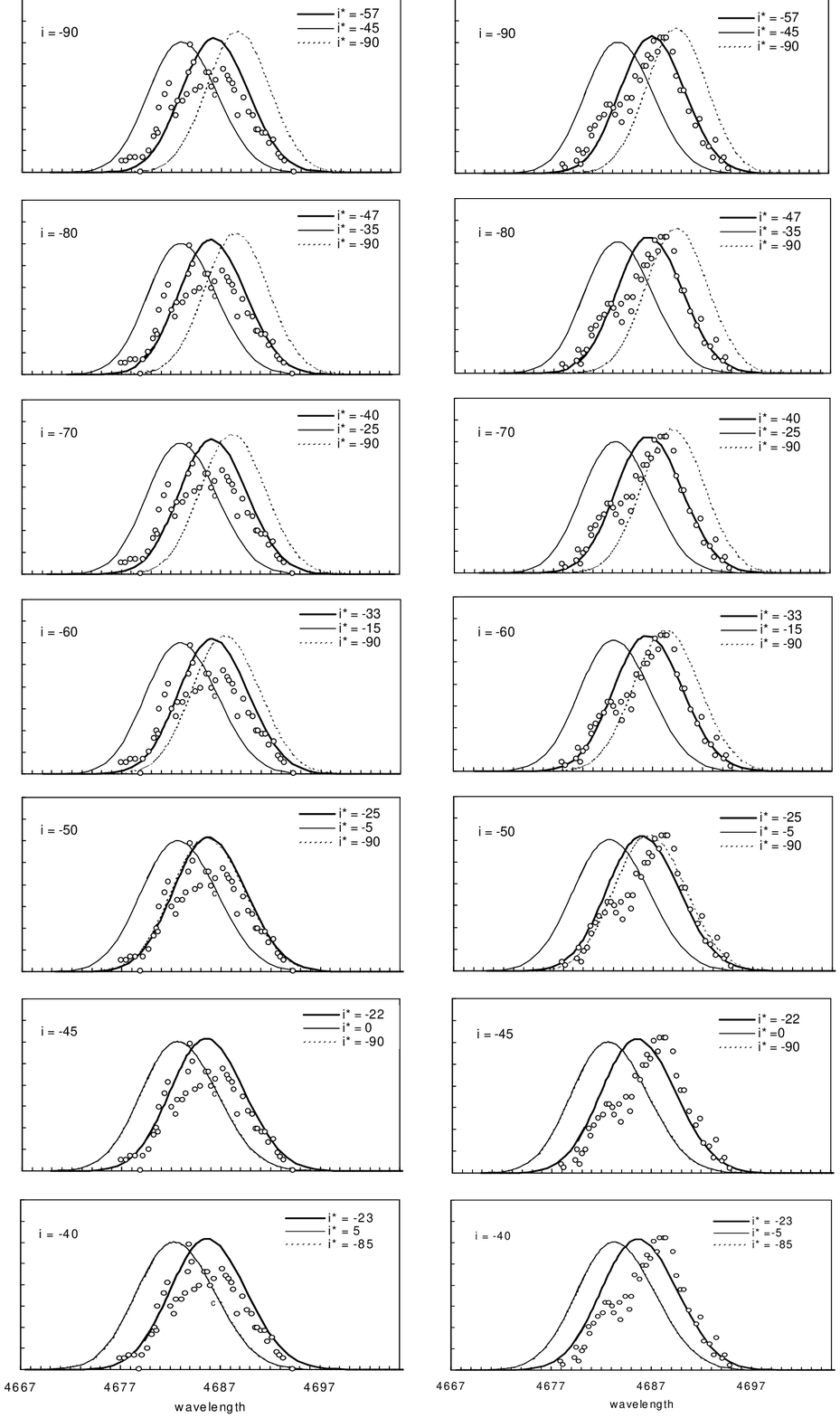}
\caption {Same as Figure 5, with $\beta = 43\degr$. Pointed line represents the minimum value of $i^*$ compatible with equation 4. }
\label{figure6}
\end{figure*}

To obtain the model profiles of the radiation reflected in the polar cap, we used the fact that the Homunculus axis forms an angle of about $45\degr$ with the line of sight (Davidson et al. 2001, Smith 2006). This angle could be obtained if the inclination $i^*$ of the Homunculus axis and the angle $\Delta \varphi^*$ between its projection into the orbital plane and the line of sight are related by:

\begin{equation}
\cos(45\degr) = \sin (i) \sin (i^*)\cos(\Delta\varphi^*)+\cos (i)\cos (i^*)
\end{equation}
 For each inclination $i$, we could find an inclination $i^*$ (and its corresponding $\Delta\varphi^*)$ that provided a good fit to the reflected profiles.  
Figure 4 shows the observations (digitalized points) and models (solid line) for the two epochs in which the VLT observations were strongest, for inclinations $ i = 90\degr$. The inclination of the Homunculus axis with respect to the orbital plane must be $59\degr$ to fit the reflected line profiles, meaning that the plane of the orbit does not coincide with the equator of the Homunculus nebula. We were also able to fit the profiles of the reflected lines for other inclinations, as can be seen in Figure 5, for $i= 90\degr, 80\degr, 70\degr, 60\degr, 50\degr, 45\degr$ and $40\degr$. The corresponding values of $i^*$ that gave the best fitting and the minim mum value of $i^*$ compatible with equation (4) are also shown in the graph. For  all inclinations $i$, the orbital plane does not coincide with the equator of the Homunculus. The value of $i^*$ that gave the best fit to the reflected data represented also the largest possible value of the inclination for which the line of sight of the Homunculus axis lie outside the contact cone; for larger inclinations, all the interior surface would be visible and the profile would present two peaks. Since this result is a consequence of the chosen aperture angle $\beta$, we constructed also line profiles for the same inclinations as in Figure 5, but for a cone aperture $\beta = 43\degr$, which corresponds to $\eta \sim 0.1$. The fitting was also satisfactory, as can be seen in Figure 6; as expected, larger values of $i^*$ were possible without resulting in a double peak profile, but the general conclusions regarding the orientation of the orbital plane with respect to the Homunculus axis did not change.

\section{Discussion}

In this paper we have shown that it is possible  to reproduce  the He\,{\sevensize II} $\lambda$4686 line profiles, mean velocities and latitude dependence obtained  by Steiner \& Damineli (2004), Stahl et al. (2005), and Martin et al. (2006) as a function of the orbital phase,  assuming that the line is formed in the shocked material that flows at both sides of the contact surface produced by  the wind-wind collision. This result  is not only important by itself, but because it validates the orbital parameters derived in Papers I and II. The main differences between the postulated orbit and other models (eg.  Damineli et al. 1997; Davidson 1997; Ishibashi et al. 1999; Smith et al. 2004) is the position of the orbital plane, which does not coincide with the equator of the Homunculus nebula, and the orientation of periastron, occurring at an angle of about $40^\circ$ from the line of sight, or 1.5 days after  conjunction for an eccentricity $e=0.95$. We will discuss below  the implications of this orbital model in  other observational features that occur during the spectroscopic events.

\begin{enumerate}
\item {{\it The minimum in the X-ray light curve.} Since, according to our model,  the secondary star is positioned in front  of $\eta$ Carinae during the spectroscopic events, for position angles $\varphi<-90\degr$ (${\rm orbital \, phase} < 0.986$), X-rays from the shocked secondary wind produced inside the conical contact surface will be seen partially absorbed by the dense cooling material concentrated in this surface.  For $\varphi>-90\degr$, part of the X-rays will be directly visible,  absorbed  only  by the secondary wind, until  the dense shocked material,  moving slowly away from the stars,  will intercept the line of sight to the observer,  producing the long lasting minimum in the X-ray light curve.\\}

\item {{\it The shell effect.} The orbital model can also explain  the light curves and spectra at optical and UV wavelengths; for  $\varphi<-90\degr$ we will detect mainly the unabsorbed   radiation from $\eta$ Carinae and its dense wind, but for  $\varphi>-90^\circ$ this radiation will start to be absorbed by the dust formed at  the contact surface, since  close to periastron passage, dust grains  can condense and grow on timescales of days and even hours (Paper I).  At the same time, the emission of the secondary star and its wind will become directly visible, explaining the dramatic change in the UV spectra   obtained with $FUSE$ satellite (Iping et al. 2005; Hillier et al. 2006).  Later,  the same dusty material that produces the X-ray fading will absorb the optical and UV radiation, giving origin to the minimum in the light curves at these wavelengths.\\}

\item{{\it The Purple Haze.} It was seen in $HST$ images as a bluish-purple glow  surrounding the central star and its origin was attributed to starlight  reflected by dust (Morse et al. 1998; Smith et al. 2000; Smith et al. 2004). The brightness distribution of the Purple Haze changed drastically during the 2003.5 spectroscopic event, with the excess UV emission seen at the east of $\eta$ Carinae before the event and at the west, afterwards.  Smith et al. (2004) suggested that this  effect is due to  the secondary star shadow  on the primary radiation. According to our interpretation, the excess UV radiation originates in the secondary star and, close to periastron, it is confined by dust absorption to the inner region of  the contact cone, which is oriented to the east before conjunction and to the west afterwards, producing the observed effect.\\}

\item{{\it Delay between the X-ray and He\,{\sevensize II} $\lambda$4686 minima in the light curves.} Martin et al. (2006) pointed out that the He\,{\sevensize II} $\lambda$4686 brightness reached its maximum  value a month after the maximum in the X-ray light curve. This is also explained by our model, since the main contribution to the X-rays comes from the hotter secondary shock region, and  absorption starts to be important when the amount of material  accumulated in the contact surface  increases, as the stars approach each other. The He\,{\sevensize II} line, on the other hand, can be formed at both sides of the contact surface, and its brightness will increase until enough material is accumulated and dust grains grow in front  of the binary system.  As  already mentioned before, both the X-rays and He\,{\sevensize II} $\lambda$4686 line will be later absorbed by dust moving in front of the binary orbit.\\} 

\item{{\it The radio emission.} According to Duncan \& White (2003), it originates in a dense, optically thick  disk. In our model the disk is coplanar with the orbit; during the spectroscopic events, a large part of the ionizing UV flux will be absorbed by the dust layer, and the ionized gas will recombine, producing the observed minima in the radio light curves and changing the shape of the 3 and 6 cm radio images (Duncan, White \& Lim 1997). During the  2003.5 spectroscopic event, the 7 mm light curve presented a sharp peak superposed to the fading flux (Abraham et al. 2005a); this excess flux was attributed to free-free emission from  dense ionized regions at both sides of the contact surface (Paper II). The emitting region was most of the time optically thick at 7 mm and therefore, its flux proportional to the area of the contact cone projected into the plane of the sky.  Since radio waves are not absorbed by dust, the emission was visible even after the spectroscopic event and the orbital parameters were determined from the shape of the light curve. In Figure 6 we compare the X-ray and 7 mm light curves close to periastron,  as well as the equivalent width of the He\,{\sevensize II} $\lambda 4686$ line obtained from Corcoran (2005), Abraham et al. (2005a), and Steiner \& Damineli (2004), respectively. We can see that the maximum in the 7 mm peak coincides with the beginning of the X-ray minimum in June 29 2003, which according to our model corresponds to the ephemerides of conjunction. The minimum in the He\,{\sevensize II} line occurred two days latter, the time necessary for dust formation.\\ }

\begin{figure}
\centering
\includegraphics[width=7cm]{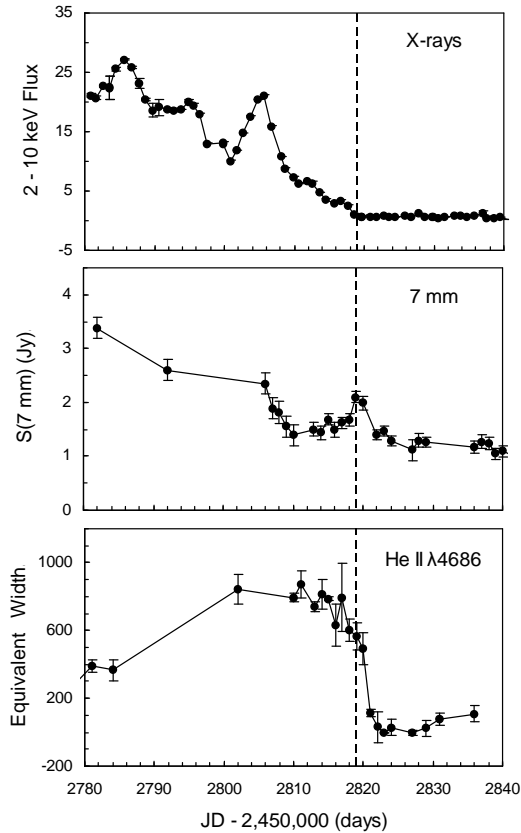}
\caption {Comparison between the X-ray light curve near the 2003.5 spectroscopic event  (indicated by a vertical pointed line) from Cororan (2005), 7 mm from Abraham et al. (2005a) and equivalent width of the He\,{\sevensize II} $\lambda$4686 line from Steiner \& Damineli (2004). }
\label{figure6}
\end{figure}

\end{enumerate}

\section{Conclusions}

In this paper we have assumed that the He\,{\sevensize II} $\lambda$4686 line, observed at orbital phases near the 2003.5 spectroscopic event by Steiner \& Damineli (2004), Stahl et al. (2005) and Martin et al. (2006), was emitted  by the shocked material that flows at both sides of the conical contact surface formed by wind-wind collision. We took into account that at these epochs, the density of the shock layer can be very large and radiative cooling very fast, allowing grain formation on timescales of days and even hours (Paper I). For that reason we assumed infinite optical depth at the contact surface, meaning that only the shock material at the observer's side of the surface contributed to the formation of the line profile. We also assumed that the flow velocity is constant and introduced turbulence with a gaussian velocity distribution. 

We constructed model profiles and compare them with the $HST$ observations reported by Martin et al. (2006) for different orbital phases, assuming a period of 2022.1 days (Steiner \& Damineli (2004) and using the orbital parameters derived in Paper II from the 7 mm light curve (eccentricity $e=0.95$, orbital inclination $i=90\degr$, epoch of conjunction June 29, 2003, angle between periastron and conjunction $\theta=40\degr)$. We found that the flow velocity and turbulence dispersion that gave the best fitting to the data were $ v_{\rm flow}\sin \beta =450$ km s$^{-1}$ and  $\sigma = 0.33 v_{\rm flow}$ km s$^{-1}$, respectively. With these parameters we were able to reproduce  the line mean velocity as a function of the orbital phase measured by Steiner \& Damineli (2004). We found that with small changes in the flow velocity and turbulence, it is also possible to obtain a good fitting to the line profiles for other inclinations and apperture angles of the contact surface cone. 

We also modeled the profiles of the He\,{\sevensize II} $\lambda$4686 line scattered by the southern  polar cap  of the Homunculus into the line of sight (Stahl et al. 2005, Martin et al. 2006). Since the model line profiles depend on the orbital phase and orientation of the observer relative to the plane of the orbit, we determined these parameters by fitting the models to the observations while fixing the angle between the observer and the axis of the Homunculus in $45\degr$.

Although the model presented here is very simple, it reproduced very well the observations and allowed us to explain other observational features,  like the ``shell like effect'', postulated to explain the optical and UV light curves (Zanella et al. 1984), the delay between  the minima at X-rays and other frequencies (Martin et al. 2006), and the variability of the ``Purple Haze'' with orbital phase (Smith et al. 2004).

\section*{Acknowledgments}

This work was partially  supported by the Brazilian agencies  
FAPESP (processes 06/57824-1 e 07/50065-0) and CNPq.

\end{document}